\documentclass[aps,preprint,tightenlines,superscriptaddress,nofootinbib,byrevtex,showpacs]{revtex4}
\usepackage{graphicx,comment,bm}
\usepackage{amssymb,latexsym}
\usepackage{amsmath,amsbsy}

\begin{document}

\unitlength=1mm

\def\a{{\alpha}}
\def\b{{\beta}}
\def\d{{\delta}}
\def\D{{\Delta}}
\def\e{{\epsilon}}
\def\g{{\gamma}}
\def\G{{\Gamma}}
\def\k{{\kappa}}
\def\l{{\lambda}}
\def\L{{\Lambda}}
\def\m{{\mu}}
\def\n{{\nu}}
\def\o{{\omega}}
\def\O{{\Omega}}
\def\S{{\Sigma}}
\def\s{{\sigma}}
\def\th{{\theta}}

\def\ol#1{{\overline{#1}}}

\def\Dslash{D\hskip-0.65em /}
\def\dslash{{\partial\hskip-0.5em /}}
\def\vslash{{\rlap \slash v}}

\def\CPT{{$\chi$PT}}
\def\QCPT{{Q$\chi$PT}}
\def\PQCPT{{PQ$\chi$PT}}
\def\tr{\text{tr}}
\def\str{\text{str}}
\def\diag{\text{diag}}
\def\order{{\mathcal O}}
\def\vit{{\it v}}
\def\vD{\vit\cdot D}
\def\am{\alpha_M}
\def\gm{\gamma_M}
\def\smb{\sigma_M}
\def\smt{\overline{\sigma}_M}
\def\tb{{\tilde b}}

\def\bea{\begin{eqnarray}}
\def\eea{\end{eqnarray}}
\def\nn{\nonumber}

\def\cS{{\mathcal S}}
\def\cC{{\mathcal C}}
\def\cB{{\mathcal B}}
\def\cT{{\mathcal T}}
\def\cQ{{\mathcal Q}}
\def\cL{{\mathcal L}}
\def\cO{{\mathcal O}}
\def\cA{{\mathcal A}}
\def\cH{{\mathcal H}}
\def\cF{{\mathcal F}}
\def\cG{{\mathcal G}}
\def\cE{{\mathcal E}}
\def\cJ{{\mathcal J}}
\def\cK{{\mathcal K}}
\def\cM{{\mathcal{M}}}
\def\fH{{\mathcal{H}}}

\def\Bbar{\overline{B}}
\def\Tbar{\overline{T}}
\def\cBbar{\overline{\cal B}}
\def\cTbar{\overline{\cal T}}
\def\cA{\mathcal A}
\def\pq{(PQ)}

\def\eqref#1{{(\ref{#1})}}

\title{Doubly Heavy Baryons and Quark-Diquark Symmetry in Quenched 
and Partially Quenched Chiral Perturbation Theory}

\author{Thomas Mehen}
\email[]{mehen@phy.duke.edu}
\affiliation{Department of Physics, Duke University,
Durham, NC 27708, USA}
\affiliation{Jefferson Laboratory, 12000 Jefferson Ave., 
Newport News, VA 23606, USA}
\author{Brian C.~Tiburzi}
\email[]{bctiburz@phy.duke.edu}
\affiliation{Department of Physics, Duke University,
Durham, NC 27708, USA}
\date{\today}

\begin{abstract} 
We extend the chiral Lagrangian with heavy quark-diquark symmetry to 
quenched and partially quenched theories. These theories are used to derive 
formulae for the chiral extrapolation of   masses and hyperfine splittings of 
double heavy baryons in lattice 
QCD simulations.  A quark-diquark symmetry prediction for the hyperfine splittings
of heavy mesons and doubly heavy baryons   is 
rather insensitive to chiral corrections in both quenched and partially quenched QCD.
Extrapolation formulae for the doubly heavy baryon electromagnetic transition moments are also 
determined for the partially quenched theory. 
\end{abstract}

\pacs{12.38.Gc, 12.39.Fe, 14.20.Lq, 12.39.Hg}
\maketitle

\section{Introduction}

Two heavy antiquarks in a doubly heavy antibaryon feel an attractive force when they are in the {\bf 3} of  color $SU(3)$ and therefore are expected
to form a compact  diquark whose size, $r_{QQ} \sim (m_Q v)^{-1}$, is much smaller than $\Lambda_{\rm QCD}^{-1}$. Here $m_Q$ is the heavy quark mass
and $v$ is the typical velocity of the antiquarks in the diquark. To the light degrees of freedom in the doubly heavy antibaryon, this diquark is
indistinguishable from a heavy quark in the $m_Q \to \infty$ limit. Therefore, the properties of ground state doubly heavy baryons should be related
to properties of singly heavy mesons. Savage and Wise~\cite{Savage:1990di} first observed that  quark-diquark symmetry  is a consequence of the heavy
quark limit and used Heavy Quark Effective Theory (HQET)~\cite{Manohar:2000dt} to derive a relation between hyperfine splittings of doubly heavy
baryons and singly heavy mesons.\footnote{The quark model has also been used to derive the relation~\cite{Lewis:2001iz,Ebert:2002ig}.} 
The correct framework for analyzing the properties of hadrons with two or more heavy quarks is  Non-Relativistic QCD
(NRQCD)~\cite{Bodwin:1994jh,Luke:1999kz,Brambilla:1999xf} and recently the heavy quark-diquark symmetry  prediction was rederived in both the potential NRQCD
\cite{Brambilla:2005yk} and velocity NRQCD \cite{Fleming:2005pd} formalisms.  An extension of  heavy hadron chiral perturbation theory
(HH$\chi$PT)~\cite{Wise:1992hn,Burdman:1992gh,Yan:1992gz} that incorporates heavy quark-diquark symmetry  was developed in Ref.~\cite{Hu:2005gf} and
used to derive a novel  quark-diquark symmetry prediction for the electromagnetic decays of the spin-3/2 members of the ground state doubly heavy
baryon doublets, as well as study  $O(1/m_Q)$ and chiral corrections to the electromagnetic decays and doubly heavy baryon masses. Some quark-diquark
symmetry predictions for the exotic heavy tetraquarks as well as a critical discussion of the applicability of heavy quark-diquark symmetry to charm
hadrons appeared in Ref.~\cite{Cohen:2006jg}.

Much of the recent theoretical work has been stimulated by the SELEX collaboration reports of candidates for doubly charm 
baryons~\cite{Moinester:2002uw,Mattson:2002vu,Ocherashvili:2004hi}. While the masses and hyperfine splittings of the observed states are consistent
with theoretical calculations in the quark model~\cite{Ebert:2002ig} as well as quenched lattice QCD, other aspects of the data are poorly understood at the
present time. For example, isospin splittings, weak decay rates,  and production yields are in disagreement with theoretical expectations. Some
excited states which would be expected to decay  electromagnetically are instead observed via their weak decay.  For a review of the  experimental
situation and discussion of theoretical interpretations of the SELEX results, see Refs.~\cite{Hu:2005gf,Cohen:2006jg, Kiselev:2002an}. Recently, the
BABAR~\cite{Aubert:2006qw} and BELLE~\cite{Chistov:2006zj} experiments searched for doubly charmed baryons  in $e^+ e^-$ collisions but failed to
observe the SELEX states. More experimental data and improved theoretical understanding of doubly charm baryons are necessary to clarify the
situation.

Lattice gauge theory studies of the properties of doubly heavy baryons with dynamical quarks would be quite useful. First principles calculation of
the spectroscopy of doubly heavy baryons could be directly compared with experiment to test our understanding of QCD and aid in the interpretation of
candidate states.  It would also be of interest to have first principle calculations of electromagnetic and weak decay matrix elements. Finally, the
lattice could  help determine  the suitability of quark-diquark symmetry for charm and bottom
hadrons. In these cases, it is possible that symmetry breaking corrections could be large enough to spoil symmetry predictions, but no systematic
study of symmetry  breaking corrections in NRQCD is yet available. A theoretical testing ground in which the heavy quark mass can be dialed at will 
could determine whether this symmetry will turn out to be useful for charm and bottom. Spectroscopic studies of doubly charm and bottom baryons in
quenched lattice QCD  appeared in Refs.~\cite{Lewis:2001iz,Mathur:2002ce,Flynn:2003vz}. The heavy quark-diquark symmetry prediction for the hyperfine
splitting holds at the 20\% level, which is reasonable given the sources of error. In QCD, corrections are expected to be  $O(v^2)$ and/or 
$O(\Lambda_{\rm QCD}/m_Q)$.  In existing lattice calculations there are also uncontrolled errors due to quenching, which, as we will see below,
could be significant. Hopefully, the near future will bring studies of doubly heavy baryons in simulations with dynamical quarks.

As a step towards such studies, it is clearly useful to develop extensions of the heavy quark-diquark symmetric version of
HH$\chi$PT~\cite{Hu:2005gf} that include lattice artifacts such as quenching and partial quenching. These theories provide 
the formulae needed for  chiral extrapolations in lattice calculations. We will focus on chiral corrections to the
doubly heavy baryon masses, especially hyperfine splittings and the quark-diquark symmetry prediction. We also apply the
formalism to the electromagnetic decays of the spin-3/2 members of the ground state doubly heavy baryon doublets.

The organization of our paper is as follows. In section~\ref{pqhhcpt}, we review the basics of quenched chiral perturbation
theory (Q$\chi$PT) and partially quenched chiral perturbation theory (PQ$\chi$PT) in the pseudoscalar 
meson sector. We then  write down versions 
of \PQCPT\ and \QCPT\ that incorporate heavy quark-diquark symmetry.  In section~\ref{mass},  we 
calculate the one-loop chiral corrections to the masses of doubly heavy baryons in \PQCPT\ and \QCPT. These expressions can be
used to perform the chiral extrapolation of lattice data on the spectrum of doubly heavy baryons provided one is in the range
of applicability of the chiral effective theory.  In section~\ref{emts}, we calculate to one loop the dipole and quadrupole transition
moments in \PQCPT. For these transitions, the lattice can be used to test the  predictions  for the electromagnetic decay
widths of the spin-3/2 members of the ground state doubly heavy baryon doublets~\cite{Hu:2005gf}. Finally, we conclude with a
brief summary in section~\ref{summy}. Some useful formuale are collected in the Appendix.

\section{\PQCPT\ and \QCPT\ for doubly heavy baryons} \label{pqhhcpt}

The formalism for constructing effective theories for quenched and partially quenched QCD
is now well established. We use the technique of graded Lie algebras
that was formulated in Refs.~\cite{Morel:1987xk,Sharpe:1992ft,Bernard:1992mk,Bernard:1994sv,Sharpe:1997by,Golterman:1998st,Sharpe:2000bc,Sharpe:2001fh}. 
Equivalently one could use the replica method~%
\cite{Damgaard:2000gh}. 
We begin by briefly reviewing the pseudoscalar meson sector of \QCPT\ and \PQCPT, and 
then write down the quenched and partially quenched chiral Lagrangians incorporating
heavy quark-diquark symmetry.  These theories are an extension of \QCPT\ and \PQCPT\
for singly heavy mesons~\cite{Sharpe:1995qp}.

\subsection{Pseudoscalar Mesons}

In the pseudoscalar meson sector, we can treat \PQCPT\ and \QCPT\ as different limits
of the same theory. This theory is described by the Lagrangian
\begin{equation}
  {\cal L} =
    \frac{f^2}{8} \str \left(
      \partial^\mu \S^\dagger \partial_\mu \S \right)
      + \frac{\l}{4}  \, \str \left( m_q \S^\dagger + m_q^\dagger \S \right)
           +\a_\Phi \partial^\mu \Phi_0 \partial_\mu \Phi_0
           - \mu_0^2 \Phi_0^2,
\label{eq:pqbosons}
\end{equation}
where the field $\Sigma$ is defined by
\begin{equation}
  \S = \exp \left( \frac{2 i \Phi}{f} \right) = \xi^2,
\end{equation}
and the meson fields appear in the $U(6|3)$ matrix,
\begin{equation}
    \Phi =
    \begin{pmatrix}
      M & \chi^\dagger\\
      \chi & \tilde M\\
    \end{pmatrix}. \label{eq:mesonmatrix}
\end{equation}
The quantities $\a_\Phi$ and $\mu_0$ are non-vanishing in the chiral limit.
The $M$ and $\tilde M$ matrices contain bosonic mesons (with quantum numbers of $q \bar{q}$ pairs and 
$\tilde{q} \bar{\tilde{q}}$ pairs, respectively, where $q$ is a valence or sea quark and $\tilde q$ is a ghost quark), 
while the $\chi$ and $\chi^\dagger$ matrices contain fermionic mesons (with quantum numbers of $\tilde q \bar{q}$
pairs and $q \bar{\tilde{q}}$ pairs, respectively).
The upper $3 \times 3$ block of the matrix $M$ contains the familiar
pions, kaons, and eta, while the remaining components consist of mesons formed
from one or two sea quarks.
The operation $\str( )$ in Eq. (\ref{eq:pqbosons}) is a supertrace over flavor indices, i.e.,
${\rm str}(A) = \sum_a \epsilon_a A_{aa}$, where $\epsilon_a = (-)^{1+\eta_a}$, and $\eta_a$ are the grading factors
of the Lie algebra.

The quark mass matrix appearing above is given by 
\begin{equation}
  m_q=\text{diag}(m_u,m_d,m_s,m_j,m_l,m_r,m_u,m_d,m_s)
.\end{equation}
We will work in the isospin limit of both the valence and sea sectors, 
where $m_d = m_u$ and $m_l = m_j$. 
Expanding the Lagrangian in Eq.~\eqref{eq:pqbosons} to lowest order in the fields,
one finds that mesons with quark content $q \ol q'$ are canonically normalized
and their masses are given by
\begin{equation}
m^2_{qq'} = \frac{\lambda}{f^2} (m_q + m_{q'}) 
\label{eq:pqmesonmass}.
\end{equation}
The propagators of the off-diagonal mesons have the usual Klein-Gordon form.

On the flavor diagonal, the situation is more complicated and 
differs dramatically between the quenched and partially quenched theories. 
The flavor singlet field that appears above is defined to be 
$\Phi_0 = {\rm str}( \Phi ) / {\sqrt 6}$.  Using this, 
the leading-order $\eta_a \eta_b$ propagator
is~\cite{Sharpe:2000bc}
\begin{eqnarray} \label{eq:neutral}
{\cal G}_{a b} (q^2) &=&
        \frac{i \epsilon_a \delta_{ab}}{q^2 - m^2_{aa} +i\epsilon} 
- \frac{i}{3} \frac{\left(q^2 - m^2_{jj} \right) \left( q^2 - m^2_{rr} \right)
	              \left( \a_\Phi q^2 - \mu_0^2 \right)}
            {\left(q^2 - m^2_{aa} +i\epsilon \right)
             \left(q^2 - m^2_{bb} +i\epsilon \right)  
             D(q^2)
             } \, .
\end{eqnarray}
The function $D(q^2)$ is defined as
\begin{equation} \notag
D(q^2) = (q^2 - m_{jj}^2)(q^2 - m_{rr}^2) + (\alpha_\Phi^2 q^2 - \mu_0^2) (q^2 - m_X^2)
,\end{equation}
and the mass $m_X$ is given by 
$m_X^2 = \frac{1}{3}\left(m^2_{jj} + 2 m^2_{rr}\right)$. 

Because the fermionic determinant in quenched QCD  is set to a constant, 
the sea quark masses are effectively infinite. The mesons containing
sea quarks decouple from the theory, leaving a $U(3|3)$ Goldstone manifold. 
In this limit, the flavor neutral propagator
in Eq.~\eqref{eq:neutral} becomes
\begin{equation}
{\cal G}^{Q}_{a b} (q^2) =
        \frac{i \epsilon_a \delta_{ab}}{q^2 - m^2_{aa} +i\epsilon} 
      - \frac{i}{3} \frac{ \a_\Phi q^2 - \mu_0^2 }
            {\left(q^2 - m^2_{aa} +i\epsilon \right)
             \left(q^2 - m^2_{bb} +i\epsilon \right)}
,\end{equation}
where we have appended a superscript $Q$ to denote that this propagator is for the quenched theory.
Notice the flavor singlet meson does not decouple as the theory still depends on the  parameters
$\a_\Phi$ and $\mu_0$. Numerical data suggest that $\a_\Phi$ is small (for example, a recent study 
finds $\alpha_\Phi = 0.03 \pm 0.03$~\cite{Bardeen:2003qz}), and we neglect it below; $\mu_0$ will 
be treated in the standard quenched power counting~\cite{Sharpe:2003hw}.

In the partially quenched theory, sea quark contributions are retained
and partially quenched QCD (PQQCD) has a strong axial anomaly. Because of this, 
the mass of the singlet field $\mu_0$ can be taken to be
on the order of the chiral symmetry breaking scale, and the $\Phi_0$ 
integrated out of the theory. The Goldstone manifold is reduced 
to $SU(6|3)$ but the resulting flavor neutral  
two-point function, however, still deviates from the familiar form in \CPT.
In PQQCD this propagator is given by the $\mu_0 \to \infty$ limit
of Eq.~\eqref{eq:neutral}, namely
\begin{equation} \label{eq:pqprop}
{\cal G}^{PQ}_{a b} (q^2) =
        \frac{i \epsilon_a \delta_{ab}}{q^2 - m^2_{aa} +i\epsilon}
        - \frac{i}{3} \frac{\left(q^2 - m^2_{jj}
            \right) \left( q^2 - m^2_{rr} \right)}
            {\left(q^2 - m^2_{aa} +i\epsilon \right)
             \left(q^2 - m^2_{bb} +i\epsilon \right)
             \left(q^2 - m^2_X +i\epsilon \right)}\, 
.\end{equation}
The \PQCPT\ flavor neutral propagator can be conveniently rewritten as
\begin{equation}
{\cal G}^{PQ}_{a b} =
         \e_a \d_{ab} P_a +
         {\cal P}_{ab}\left(P_a,P_b,P_X\right),
\end{equation}
where
\begin{eqnarray}
     P_a &=& \frac{i}{q^2 - m^2_{aa} +i\e},\ 
     P_b = \frac{i}{q^2 - m^2_{bb} +i\e},\ 
     P_X = \frac{i}{q^2 - m^2_X +i\e}, \,
\nonumber\\
     {\cal P}_{ab}\left(A,B,C\right) &=& 
           -\frac{1}{3}\left[
             \frac{\left( m^2_{aa} - m^2_{jj}\right)
                   \left( m^2_{aa} - m^2_{rr}\right)}
                  {\left( m^2_{aa} - m^2_{bb}\right)
                   \left( m^2_{aa} - m^2_X\right)}
                 A
            +\frac{\left( m^2_{bb} - m^2_{jj}\right)
                   \left( m^2_{bb} - m^2_{rr}\right)}
                  {\left( m^2_{bb} - m^2_{aa}\right)
                   \left( m^2_{bb} - m^2_X\right)}
                 B \right.\, 
\nonumber\\
&&\qquad\quad\left.
            +\frac{\left( m^2_X - m^2_{jj}\right)
                   \left( m^2_X - m^2_{rr}\right)}
                  {\left( m^2_X - m^2_{aa}\right)
                   \left( m^2_X - m^2_{bb}\right)}
                 C\ \right].
\label{eq:Pab}
\end{eqnarray}

\subsection{Doubly heavy baryons}

As discussed earlier, in the heavy quark limit the properties of singly heavy mesons and doubly heavy baryons
are related by a $U(5)$ quark-diquark symmetry~\cite{Savage:1990di}.
Recently a chiral Lagrangian incorporating this $U(5)$ symmetry has been derived~\cite{Hu:2005gf}. 
It is written in terms of a super-field $\fH$ given by
\begin{equation} \label{eq:H}
\fH_{a,\mu\b} = H_{a,\a \b} + T_{a,i\b} \, 
,\end{equation}
where $a$ is an anti-fundamental flavor index, $\a$ and $\b$ are Pauli spinor indices, while $i$ is a vector spin index. 
The super-spin index $\mu = 1$--$5$ and is related to $\a$ and $i$ via 
\begin{equation}
\mu = \a (\d_{\mu}^1 + \d_{\mu}^2) + (i + 2) (\d_{\mu}^3 + \d_{\mu}^4 + \d_{\mu}^ 5)
.\end{equation}
The field $H_{a,\a \b}$ is the heavy meson field of HH$\chi$PT \cite{Wise:1992hn,Burdman:1992gh,Yan:1992gz} 
 in the rest frame of the heavy meson:
\begin{equation} \label{eq:Hrest}
H_{a,\a \b} = \bm{P}^*_a \cdot \bm{\sigma}_{\a \b} + P_a \d_{\a\b} 
.\end{equation}
$H_a$ contains the vector $(\bm{P}^*)$ and pseudoscalar $(P)$ field components. 
The field $T_{a,i\b}$ is the doubly heavy baryon field,
\begin{equation} \label{eq:T}
T_{a,i\b} = \sqrt{2} \left( \Xi^*_{a,i\b} + \frac{1}{\sqrt{3}} \Xi_{a,\gamma} \sigma^i_{\gamma\b} \right)
,\end{equation}
that contains the spin-$1/2$ $(\Xi)$ and spin-$3/2$ $(\Xi^*)$ field components. The spin-$3/2$ field satisfies
a non-relativistic Rarita-Schwinger constraint, $\Xi^*_{a,i\b} \sigma^i_{\b \gamma} = 0$.

In terms of the super-field $\fH$, the Lagrangian for the quenched (Q) 
theory is 
\begin{eqnarray} \label{eq:LQ}
\cL^{Q} &=& 
\left( \fH^\dagger (\fH i \overset{\leftarrow}{D}_0 )\right) 
- 
g^{Q} \left( \fH^\dagger \fH \bm A \cdot \bm{\sigma} \right)
- 
g'{}^{Q} \left( \fH^\dagger \fH \bm{\sigma} \right) \cdot \str (\bm{A})
\notag \\
&& + 
\frac{\D^{Q}_H}{4} \left( \fH^\dagger \bm{\Sigma} \cdot \fH \bm{\sigma} \right)
+ 
\sigma^{Q} \left( \fH^\dagger \fH \cM \right)
+
\sigma'{}^{Q} \left(\fH^\dagger \fH \right) \str (\cM)
,\end{eqnarray} 
while that for the partially quenched theory (PQ) is
\begin{eqnarray} \label{eq:LPQ}
\cL^{PQ} &=& 
\left( \fH^\dagger (\fH i \overset{\leftarrow}{D}_0 )\right) 
- 
g^{PQ} \left( \fH^\dagger \fH \bm A \cdot \bm{\sigma} \right)
\notag \\
&& + 
\frac{\D^{PQ}_H}{4} \left( \fH^\dagger \bm{\Sigma} \cdot \fH \bm{\sigma} \right)
+ 
\sigma^{PQ} \left( \fH^\dagger \fH \cM \right)
+
\sigma'{}^{PQ} \left(\fH^\dagger \fH \right) \str (\cM)
.\end{eqnarray} 
Both Lagrangians include the leading quark-diquark symmetry breaking term. 
The bracket $()$ notation denotes the contraction of flavor, spinor and super-spin indices of field bilinears. 
For a generic super-spin matrix $X$, flavor matrix $Y$, and spin matrix  $Z$, $()$ denotes
\begin{equation}
\left( \fH^\dagger X \, \fH \, Y \, Z \right)
= 
\sum_{\mu, \nu} \sum_{\a, \b} \sum_{a, b}  
\epsilon_{a} 
\fH^\dagger_{a,\mu \a} 
X_{\mu \nu} 
\fH_{b,\nu \b} 
Y_{ba} Z_{\b\a}
.\end{equation} 
In Eqs.~(\ref{eq:LQ}) and (\ref{eq:LPQ}),  
$\bm{A}$ is the spatial part of the axial-vector pion field,
$A_\mu = \frac{i}{2} \left( \xi \partial_\mu \xi^\dagger - \xi^\dagger \partial_\mu \xi \right)$,
and $\cM$ 
is the mass operator defined by 
$\cM = \frac{1}{2} \left( \xi m_q \xi + \xi^\dagger m_q \xi^\dagger \right)$.  
The 
$U(5)$ 
super-spin matrix 
$\bm{\Sigma}$ 
is given by
\begin{equation}
\bm{\Sigma}_{\mu \nu} = 
\begin{pmatrix}
\bm{\sigma}_{\a\b} & 0             \\
                0  & \bm{\cT}_{jk} \\ 
\end{pmatrix}
,\end{equation}
where 
$(\cT^i)_{jk} = - i \epsilon_{ijk}$.

Notice that there are separate Lagrangians for  quenched and partially quenched QCD. 
The low-energy constants of the two theories are different, 
e.g. 
$g^{PQ} \neq g^{Q}$. 
Notice also that in \QCPT\ the flavor singlet field cannot be integrated out. 
Consequently 
$\str (\bm{A}) \neq 0$ 
and the axial hairpin coupling 
$g'{}^{Q}$ 
remains. Additionally in \QCPT,
$\str (\cM) = 0$ 
to leading order, and so we can effectively set  
$\sigma'^{Q} = 0$ 
for our calculations. 
There is no sector of \QCPT\ that contains \CPT, so in general quenched observables are unrelated to QCD observables. 
The sea sector of PQQCD is QCD, and hence \PQCPT\ contains the low-energy constants of \CPT. Restricting the 
flavor indices to the sea sector in Eq.~\eqref{eq:LPQ}, we recover the chiral Lagrangian of Ref.~\cite{Hu:2005gf}, 
so  $g^{PQ}$, $\D_H^{PQ}$, $\sigma^{PQ}$ and $\sigma'^{PQ}$ are identical to the low-energy constants
$g$, $\D_H$, $\sigma$, and $\sigma'$ of \CPT, respectively. At this order there are no additional operators in 
the partially quenched Lagrangian. 
Carrying out the super-spin sums in Eqs.~\eqref{eq:LQ} and \eqref{eq:LPQ}, 
we produce separate terms for heavy mesons and doubly heavy 
baryons but with the same low-energy constants. 
The heavy meson sector of the \PQCPT\ and \QCPT\ Lagrangians above is that of Ref.~\cite{Sharpe:1995qp}.

\section{Masses} \label{mass}

In this section, we use the quenched and partially quenched chiral 
Lagrangians that incorporate quark-diquark symmetry to calculate 
chiral corrections to the masses of doubly heavy baryons. 
The results can be used for the chiral 
extrapolation of lattice data. \PQCPT\ is considered in subsection III.A and \QCPT\ is discussed in subsection III.B.
 It is important to keep in mind that while \PQCPT\ actually 
 contains \CPT, there is no such connection between \QCPT\ and \CPT.

Near the chiral limit, the masses of doubly heavy baryons generically
have the form
\begin{eqnarray}
M_{\Xi} 
&=& 
M_0 - \frac{1}{2} \Delta_H +  M_{\Xi}^{(1)} + M_{\Xi}^{(3/2)} + \ldots
\notag \\
M_{\Xi^*} 
&=& 
M_0 + \frac{1}{4} \Delta_H + M_{\Xi^*}^{(1)} + M_{\Xi^*}^{(3/2)} + \ldots
\notag
,\end{eqnarray}
where $M_0$ is the mass of the doubly heavy baryons in the chiral 
and heavy quark limit and $\Delta_H$ is the hyperfine mass splitting of the heavy mesons
in the chiral limit.\footnote{%
Strictly speaking $\Delta_H$ is not same the parameter appearing 
in the Lagrangian above, Eq.~\eqref{eq:LPQ}, which in this footnote
we denote by $\Delta_H^{0}$. This parameter, which transforms as a chiral 
singlet, is corrected by a string of operators that differ only by scaling 
with powers of $\Delta_{H}^{0} / \Lambda_\chi$. Thus we subsume
this dependence as $\Delta_{H} = \Delta_H(\Delta^0_H)$ including also 
contributions to $\Delta_H$ arising from loop graphs. 
In this way we obtain $\Delta_H$ as the chiral limit value.
The same is true of all other low-energy constants in Eq.~\eqref{eq:LPQ}.
We similarly drop their polynomial dependence on $\Delta_{H}^{0} / \Lambda_\chi$, 
and absorb contributions from loop graphs to work with the values of the low-energy 
constants in the chiral limit.
For more details see, e.g.~Ref~\cite{Tiburzi:2004kd}. 
}
The hyperfine splitting of the $\Xi^*$ and $\Xi$ baryons is $\frac{3}{4}\Delta_H$ in the chiral limit
due to quark-diquark symmetry.
The term $M^{(1)}$ is proportional to $m_q$ and comes from tree-level counterterms.
The term $M^{(3/2)}$ comes from the one-loop diagrams and depends on $m$, the Goldstone boson mass,
and $\delta$, the mass splitting between the external  and virtual doubly heavy baryons.
Goldstone boson masses are $\propto \sqrt{m_q}$, while $\delta$ receives contributions from
$SU(3)$ breaking counterterms, which are $\propto m_q$, and/or hyperfine splittings, 
which  scale as $1/m_Q$. In our power counting we take $\delta \sim m \sim Q$ in which case 
$M^{(3/2)} \propto Q^3$. In the chiral and heavy quark limits,  $\delta$ can be neglected relative to 
$m$ and then $M^{(3/2)} \propto m_q^{3/2}$. The ellipsis denotes higher-order terms in the $Q$ expansion.

\subsection{\PQCPT}

At linear order in the quark mass, the contributions to the masses
of doubly heavy baryons at tree level come from the mass operators
in Eq.~\eqref{eq:LPQ}:
\begin{equation}
M_{\Xi^*_a}^{(1)} = M_{\Xi_a}^{(1)} = - \sigma (m_q)_a - \sigma' \str (m_q) 
,\end{equation}
where $\str (m_q) = 2 m_j + m_r$ in the isospin limit, and the subscript $a$ denotes the 
light-quark anti-fundamental index of the doubly heavy baryon. 
Combining these
results with the mass splitting in the chiral limit, we then define the baryon
mass splittings
\begin{eqnarray}
\D_{ba}   &=&   M_{\Xi_b} - M_{\Xi_a} =  M_{\Xi_b^*} - M_{\Xi_a^*} = \sigma [ (m_q)_a - (m_q)_b ],
\\
\D_{ba}^* &=&   M_{\Xi_b^*} - M_{\Xi_a} = \frac{3}{4} \D_H  + \sigma [ (m_q)_a - (m_q)_b ]
.\end{eqnarray}

\begin{figure}[tb]
  \centering
  \includegraphics[width=0.5\textwidth]{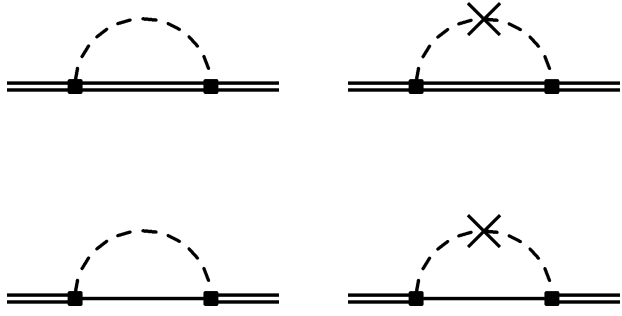}%
  \caption{
    One-loop graphs which give contributions to the masses of the doubly heavy baryons in \PQCPT.  
    The single and double lines correspond to $\Xi$ baryons or $\Xi^*$ baryons, respectively, while the 
    dashed lines correspond to mesons.  The filled squares denote the non-singlet 
    axial coupling given in Eq.~(\ref{eq:LPQ}), while crosses denote the hairpin interaction. 
  }
  \label{fig:NLOT}
\end{figure}

At one-loop order, the doubly heavy baryons receive mass contributions
in \PQCPT\ from the sunset diagrams shown in Fig.~\ref{fig:NLOT}. 
The vertices in these diagrams are generated from the pion-baryon interaction terms in Eq.~\eqref{eq:LPQ}, 
while the hairpin interaction corresponds to the disconnected part of the flavor neutral propagator in Eq.~\eqref{eq:pqprop}.  
Calculating these \PQCPT\ loop diagrams, we find
\begin{eqnarray}
M_{\Xi^*_a}^{(3/2)} 
&=& 
\frac{g^2}{(4 \pi f)^2} 
\sum_{b = j,l,r}
\Bigg[ 
\frac{5}{9} K(\D_{ba}, m_{ba}, \mu) 
+ 
\frac{4}{9} K(-\D_{ab}^*, m_{ba}, \mu) 
\Bigg]
\notag \\
&& + 
\frac{g^2}{(4 \pi f)^2}
\Bigg[
\frac{5}{9} \cK (0 ,m_{aa}, m_{aa}, \mu)
+ 
\frac{4}{9} \cK (-\D_{aa}^*, m_{aa}, m_{aa}, \mu)
\Bigg]
,\end{eqnarray}
for the spin-$3/2$ doubly heavy baryons, $\Xi^*_a$, and
\begin{eqnarray}
M_{\Xi_a}^{(3/2)} 
&=& 
\frac{g^2}{(4 \pi f)^2} 
\sum_{b = j,l,r}
\Bigg[ 
\frac{1}{9} K(\D_{ba}, m_{ba}, \mu) 
+ 
\frac{8}{9} K(\D_{ba}^*, m_{ba}, \mu) 
\Bigg]
\notag \\
&& + 
\frac{g^2}{(4 \pi f)^2}
\Bigg[
\frac{1}{9} \cK (0 ,m_{aa}, m_{aa}, \mu)
+ 
\frac{8}{9} \cK (\D_{aa}^*, m_{aa}, m_{aa}, \mu)
\Bigg]
,\end{eqnarray}
for the spin-$1/2$ doubly heavy baryons, $\Xi_a$. 
The non-analytic function $K(\d,m,\mu)$  is 
\begin{equation}
K(\d,m,\mu) 
= 
2 (\d^2 - m^2) 
\left[ 
m R \left( \frac{\d}{m} \right)
- 
\d \log \left( \frac{m^2}{\mu^2} \right)
\right] 
+ \d \, m^2 \log \left( \frac{m^2}{\mu^2} \right)
,\end{equation}
where  $R(x)$ is defined by 
\begin{equation} \label{eq:R}
R(x) = 
\sqrt{x^2 - 1} 
\log \left( \frac{x - \sqrt{x^2 - 1 + i \epsilon}}{x + \sqrt{x^2 - 1 + i \epsilon}} \right)
,\end{equation}
while the function 
$\mathcal{K}(\d,m_{aa}, m_{bb}, \mu)$ 
which arises from hairpins is given by
\begin{equation}
\mathcal{K}(\d, m_{aa}, m_{bb}, \mu) 
= 
\mathcal{P}_{ab} 
\Big[ 
K(\d, m_{aa}, \mu), 
K(\d, m_{bb},\mu),
K(\d, m_X, \mu)
\Big]
,\end{equation}
with 
$\mathcal{P}_{ab}$ 
appearing in Eq.~\eqref{eq:Pab}, and a limit is understood when 
$b = a$.

\begin{figure}
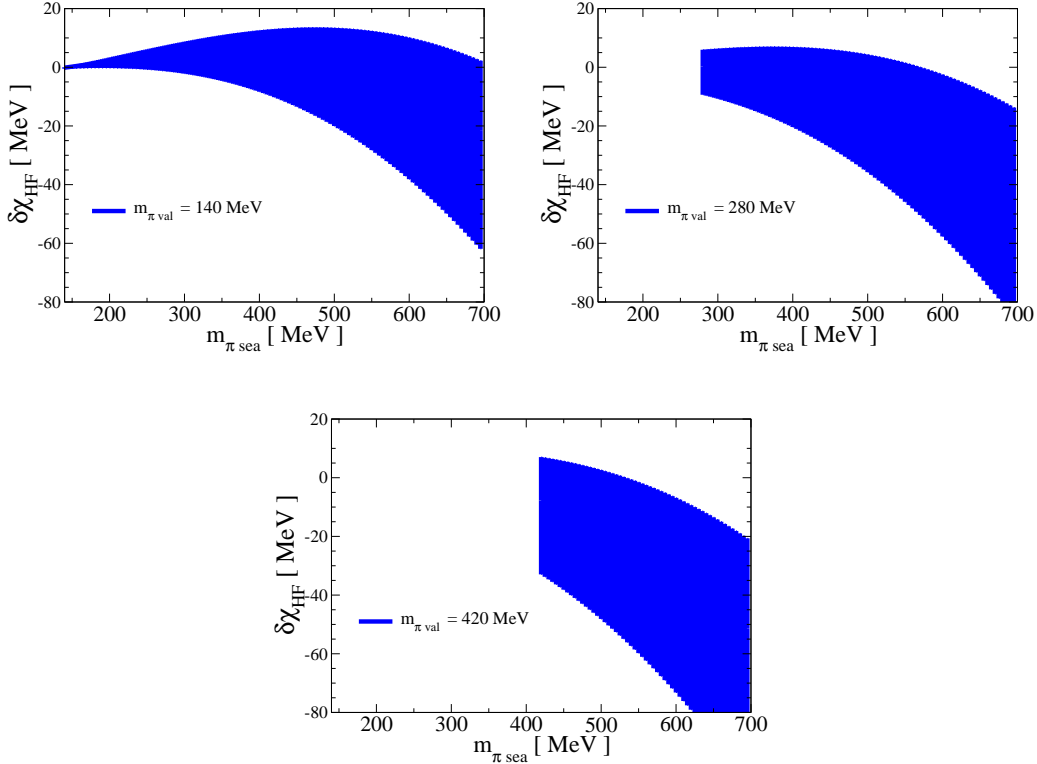

  \centering
  \includegraphics[width=0.4\textwidth]{splitPQ1.eps} \hspace{0.1 in}
  \includegraphics[width=0.4\textwidth]{splitPQ2.eps}  \\ 
  \vspace{0.3 in}
  \includegraphics[width=0.4\textwidth]{splitPQ3.eps}%
  \caption{$\delta \chi_{HF}$ as a function of  $m_{\pi \, {\rm sea}}$ for different values of   $m_{\pi \, {\rm val}}$.
  The width of the bands is the result of varying $\mu$ between 500 MeV and 1500 MeV.
   }
  \label{fig:splitPQ}
\end{figure}

To study the chiral non-analytic behavior of doubly heavy baryon mass splittings in \PQCPT, 
we need values for the low-energy constants. We choose 
$g = 0.6$~\cite{Fajfer:2006hi}, $\D_H = 140 \, \texttt{MeV}$, 
and 
$f = 130 \, \texttt{MeV}$.
The $SU(3)$ splitting of the ground state $D$ mesons is 
$\approx 100 \, \texttt{MeV}$. Assuming the same $SU(3)$ splitting for the doubly charm baryons
we obtain $\sigma /\l = - (337 \, \texttt{MeV})^{-3}$. 
We shall take the strange sea quark to be 
degenerate with its valence counterpart,
$m_r = m_s$, 
and fix the value at the physical
strange quark mass. This leaves three parameters
that can be varied, the valence pion mass, $m_{\pi \, {\rm val}} = m_{uu}$, 
the sea pion mass,  $m_{\pi \, {\rm sea}} = m_{jj}$, and the renormalization scale, 
$\mu$. 
While the 
$\mu$ 
dependence in the above expressions is exactly canceled by 
$\mu$-%
dependent counterterms at this order, we have not included these counterterms,
nor do we have a reliable way to estimate them. Hence we vary 
$\mu$ 
to effectively obtain a range for the unknown counterterm contributions.

In Fig.~\ref{fig:splitPQ}, we assess the non-analytic chiral corrections
to the hyperfine splitting of non-strange $\Xi^*$ and $\Xi$ baryons
in PQ$\chi$PT. To see the effect of partial quenching on this observable, we 
plot the difference between the  chiral non-analytic contribution to the hyperfine
splitting in PQ$\chi$PT  and the corresponding non-analytic correction
in $\chi$PT:
\begin{equation} \label{eq:delta}
\d \chi_{HF}  = \left( M_{\Xi^*}^{(3/2)} - M_\Xi^{(3/2)} \right)_{{\rm PQ}\chi{\rm PT}}
- \left( M_{\Xi^*}^{(3/2)} - M_\Xi^{(3/2)} \right)_{\chi{\rm PT}} \, ,
\end{equation}
as a function of 
$m_{\pi \, \text{sea}}$ 
for a few different values of  
$m_{\pi \, \text{val}}$.
For each value of $m_{\pi \, {\rm val}}$, we let   $m_{\pi \, {\rm sea}}$ range from  $m_{\pi \, {\rm val}}$
up to the mass of eta-strange, 
$m_{\eta_s} = m_{ss} = m_{rr} \approx 700 \, \texttt{MeV}$. 
Thus the smallest value of   $m_{\pi \, {\rm sea}}$ corresponds to an unquenched theory at 
$m_\pi = m_{\pi \, \text{val}}$, 
while the largest value of  $m_{\pi \, {\rm sea}}$ corresponds to a completely degenerate 
sea at the strange quark mass. The bands 
correspond to varying 
$\mu$ 
from 
$500 \, \texttt{MeV}$ 
to 
$1500 \, \texttt{MeV}$, which is chosen to be the same 
for both $\chi$PT and PQ$\chi$PT. For the $\chi$PT hyperfine splitting, we use the physical Goldstone boson
masses, so by construction $\d \chi_{HF} = 0 $ when $m_{\pi \, \text{sea}} = m_{\pi \, \text{val}} = m_\pi$.
As demonstrated by Fig.~\ref{fig:splitPQ}, chiral non-analytic corrections to the hyperfine splitting 
are sizable, and increase with increasing  $m_{\pi \, {\rm sea}}$. Additionally
corrections are enhanced as  $m_{\pi \, {\rm val}}$ increases. 
While we do not trust the chiral expansion at large pion masses, we have
kept these in the plot to show that chiral corrections can become 
large and tend to push the hyperfine splitting down by tens of $\texttt{MeV}$.

\begin{figure}
  \centering
  \includegraphics[width=0.55\textwidth]{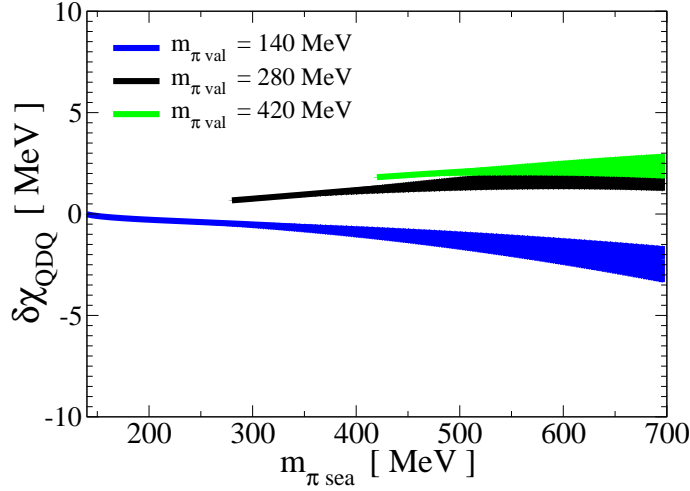}%
  \caption{$\delta \chi_{QDQ}$ as a function of  $m_{\pi \, {\rm sea}}$ for different values of  $m_{\pi \, {\rm val}}$.
  The width of the bands is the result of varying $\mu$ between 500 MeV and 1500 MeV.
   }
  \label{fig:deltaPQ}
\end{figure}

We can use our results to check the chiral corrections to the mass splitting relation 
predicted by heavy quark-diquark symmetry~\cite{Savage:1990di,Fleming:2005pd,Brambilla:2005yk}
\bea 
M_{\Xi^*}- M_{\Xi} = \frac{3}{4}(M_{P^*} - M_P) \, .
\eea
In Fig.~\ref{fig:deltaPQ}, we plot 
the difference between the nonanalytic chiral corrections to this relation
in PQ$\chi$PT and $\chi$PT, defined by
 \bea \label{eq:Delta}
\d \chi_{QDQ}  
&=& 
\left[M_{\Xi^*}^{(3/2)} - M_\Xi^{(3/2)} -  \frac{3}{4} 
\left( M_{P^*}^{(3/2)} - M_{P}^{(3/2)} \right)\right]_{{\rm PQ}\chi{\rm PT}} \nn \\
&&-\left[M_{\Xi^*}^{(3/2)} - M_\Xi^{(3/2)} -  \frac{3}{4} 
\left( M_{P^*}^{(3/2)} - M_{P}^{(3/2)} \right)\right]_{\chi{\rm PT}}
,\eea
in the non-strange sector.
Here $M_{P^*}^{(3/2)}$ and $M_{P}^{(3/2)}$ are the chiral contributions
to the masses of the vector and pseudoscalar heavy mesons, 
respectively~\cite{Sharpe:1995qp}. As in Fig.~\ref{fig:splitPQ}, we  vary  $\mu$ from $500 \, \texttt{MeV}$ 
to $1500 \, \texttt{MeV}$ for three different values of $m_{\pi \, {\rm val}}$.
Fig.~\ref{fig:deltaPQ} shows that the chiral corrections to the heavy quark-diquark symmetry prediction
are remarkably small compared  to chiral corrections to the hyperfine splittings and much less sensitive to $\mu$.
It was demonstrated in Ref.~\cite{Hu:2005gf} that chiral corrections 
to the quark-diquark symmetry prediction for the hyperfine splittings are small
and insensitive to $\mu$.  Interestingly,  this conclusion 
is not changed as the pion mass is increased, nor is it altered by 
partial quenching. 
While the hyperfine splittings of heavy mesons and doubly heavy baryons
receive considerable chiral corrections for larger pion masses, 
the quark-diquark symmetry relation fortuitously does not. Lattice calculations of this 
quantity will not suffer large uncertainty due to chiral extrapolation.

\subsection{\QCPT}

\begin{figure}
  \centering
  \includegraphics[width=0.5\textwidth]{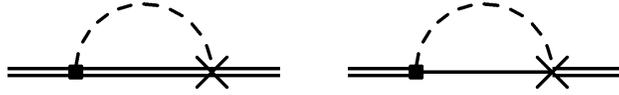}%

  \caption{Additional loop graphs which contribute to the masses of the doubly heavy baryons in \QCPT.  
    The single and double lines correspond to $\Xi$ baryons or $\Xi^*$ baryons, respectively, while the 
    dashed lines correspond to mesons.  The filled squares denote the non-singlet 
    axial coupling, and the cross denotes the axial hairpin interaction given in Eq.~\eqref{eq:LQ}. 
  }
  \label{fig:NLOTQ}
\end{figure}

In the quenched theory we can similarly calculate the masses of doubly heavy baryons to one-loop order.
The tree level contribution from $SU(3)$ breaking counterterms is 
\begin{equation}
M_{\Xi^*_a}^{(1)} = M_{\Xi_a}^{(1)} = - \sigma^Q (m_q)_a 
.\end{equation}
The tree-level baryon mass splittings are
\begin{eqnarray}
\D_{ba}   &=&   M_{\Xi_b} - M_{\Xi_a} =  M_{\Xi_b^*} - M_{\Xi_a^*} = \sigma^Q [ (m_q)_a - (m_q)_b ],
\\
\D_{ba}^* &=&   M_{\Xi_b^*} - M_{\Xi_a} = \frac{3}{4} \D_H^Q  + \sigma^Q [ (m_q)_a - (m_q)_b ]
.\end{eqnarray}

The differences in low-energy constants notwithstanding,
the one-loop mass contributions in \QCPT\ are considerably altered from 
those in \PQCPT\ and \CPT. The sunset diagrams in Fig.~\ref{fig:NLOT}
again contribute, however, 
it is clear that only hairpin contributions survive quenching. 
There are additional one-loop diagrams present in the quenched theory 
that have no counterparts in \PQCPT\ or \CPT. These diagrams are depicted in 
Fig.~\ref{fig:NLOTQ} and involve the axial hairpin interaction in Eq.~\eqref{eq:LQ}.  
Combining the hairpin diagrams and axial hairpin diagrams, we arrive at the 
one-loop quenched mass contributions
\begin{eqnarray}
M_{\Xi^*_a}^{(3/2)} 
&=& 
\frac{2 g^Q g'{}^Q}{(4 \pi f)^2}
\Bigg[
\frac{5}{9} K (0 ,m_{aa}, \mu)
+ 
\frac{4}{9} K (-\D_{aa}^*, m_{aa}, \mu)
\Bigg]
\notag \\
&& + 
\frac{(g^Q)^2 \mu_0^2}{3 (4 \pi f)^2} 
\, \frac{\partial}{\partial m_{aa}^2}
\Bigg[ 
\frac{5}{9} K(0, m_{aa}, \mu) 
+ 
\frac{4}{9} K(-\D_{aa}^*, m_{aa}, \mu) 
\Bigg]
,\end{eqnarray}
for the spin-$3/2$ doubly heavy baryons, and
\begin{eqnarray}
M_{\Xi_a}^{(3/2)} 
&=& 
\frac{2 g^Q g'{}^Q}{(4 \pi f)^2}
\Bigg[ 
\frac{1}{9} K(0, m_{aa}, \mu) 
+ 
\frac{8}{9} K(\D_{aa}^*, m_{aa}, \mu) 
\Bigg]
\notag \\
&& + 
\frac{(g^Q)^2 \mu_0^2}{3 (4 \pi f)^2} 
\, \frac{\partial}{\partial m_{aa}^2}
\Bigg[
\frac{1}{9} K (0 ,m_{aa}, \mu)
+ 
\frac{8}{9} K (\D_{aa}^*, m_{aa}, \mu)
\Bigg]
,\end{eqnarray}
for the spin-$1/2$ doubly heavy baryons.

\begin{figure}
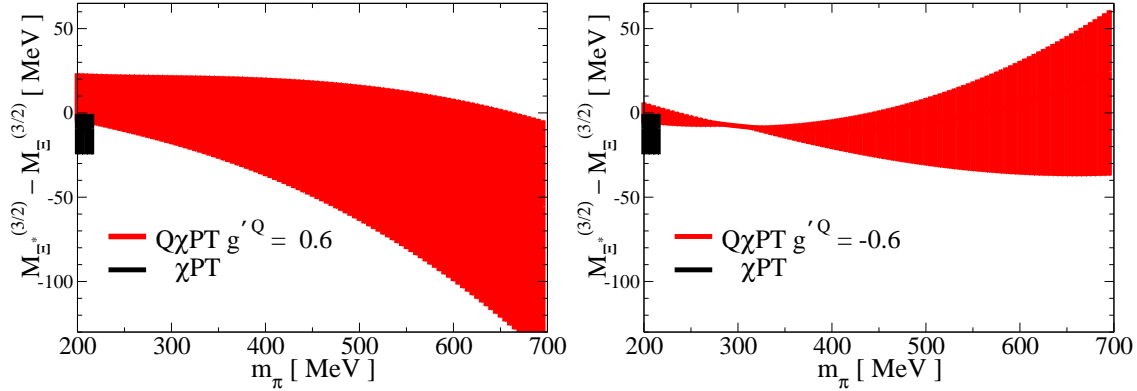

  \centering
  \includegraphics[width=0.45\textwidth]{splitQ1.eps}
  \includegraphics[width=0.45\textwidth]{splitQ2.eps}%
  \caption{
   The quenched chiral correction to the hyperfine splitting, $M^{(3/2)}_{\Xi^*} - M^{(3/2)}_\Xi$,  
  as a function of the quenched pion mass.   
   The \CPT\ band shows the corresponding range of chiral corrections to the hyperfine splitting in QCD.
}
  \label{fig:splitQ}
\end{figure}

\begin{figure}
  \centering
   \includegraphics[width=0.55\textwidth]{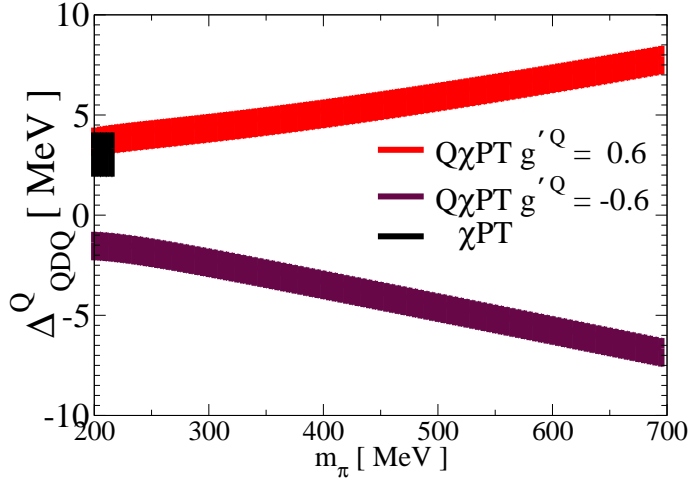}%
  \caption{The quenched chiral correction to the heavy quark-diquark symmetry breaking parameter, $\Delta^Q {}_{QDQ}$,  
	   as a function of the quenched pion mass.  The \CPT\ band shows the
           corresponding range of chiral corrections to quark-diquark symmetry in QCD.
   }   
  \label{fig:deltaQ}
\end{figure}

With the quenched chiral non-analytic corrections in hand, we can investigate
their impact on the spectrum of doubly heavy baryons as we did for
the partially quenched theory.
To determine these corrections, we must estimate the 
parameters in the quenched chiral Lagrangian. Little is known about these parameters. 
We adopt a central value of $\mu_0 = 700 \, \texttt{MeV}$ from the analysis of Ref.~\cite{Bardeen:2003qz}. 
The remaining parameters are taken to be the same values as in \CPT, though 
there is no justification for this. 
Finally we use two values, $g'^Q = \pm 0.6$, for the parameter that has no analogue in the unquenched theory. 
In Fig.~\ref{fig:splitQ}, we plot the quenched chiral
correction to the hyperfine splitting, $M^{(3/2)}_{\Xi^*} - M^{(3/2)}_\Xi$, 
as a function of the pion mass. The band plotted corresponds to 
a variation of $\mu$ from $500 \, \texttt{MeV}$ to $1500\, \texttt{MeV}$. 
As in the partially quenched theory, we see that the mass splitting receives 
sizable chiral corrections that increase with the pion mass. 
The trend is the same as in the partially quenched theory provided $g^Q$ and $g'^Q$ have the
same sign:
quenched lattice calculations at larger pion masses than that in nature will 
underestimate the hyperfine splitting. 
When $g^Q$ and $g'^Q$ have opposite signs, it is possible for the splitting to 
be overestimated. 
Lastly we can test how well heavy quark-diquark symmetry holds up in the
presence of quenched chiral corrections. In Fig.~\ref{fig:deltaQ}, we 
plot the quenched chiral correction to the quark-diquark symmetry relation for the hyperfine splittings,
\bea
\Delta^Q {}_{QDQ} =  M^{(3/2)}_{\Xi^*} - M^{(3/2)}_\Xi - \frac{3}{4}(M^{(3/2)}_{P^*} - M^{(3/2)}_{P}) \, ,
\eea
as a function of the pion mass, again varying $\mu$ from 500 to 1500 MeV. Surprisingly, the quenched 
chiral corrections do not sizably alter the quark-diquark symmetry prediction. The corrections are only a few $\texttt{MeV}$. 
The $\mu$ dependence of the result is minimal and in fact one can show analytically that the $\mu$
dependence vanishes for this 
observable if $g^{\prime Q}=0$.

\section{Electromagnetic transitions in \PQCPT} \label{emts}

\begin{figure}
  \centering
  \includegraphics[width=0.5\textwidth]{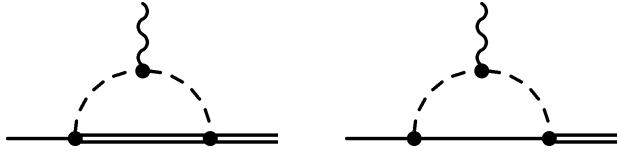}%
  \caption{
    One-loop graphs contributing to the doubly heavy baryon electromagnetic transitions in \PQCPT.  
    The single lines correspond to $\Xi$ baryons, double lines to $\Xi^*$ baryons.
    The dashed lines correspond to mesons, while the wiggly lines correspond to photons.  
    The coupling of Goldstone mesons to baryons is the non-singlet axial coupling given in Eq.~(\ref{eq:LPQ}).
  }
  \label{fig:NLOtrans}
\end{figure}

As a final application we determine the electromagnetic transition moments of the doubly 
heavy baryons in \PQCPT\ and \CPT. To include electromagnetism into the theory, we merely 
gauge a background $U(1)$ field. In doing so, recall
that the light-quark electric charge matrix $\cQ$ is not uniquely
defined in PQQCD~\cite{Golterman:2001qj}.  
The only constraint one must impose is that
the charge matrix $\cQ$ has vanishing
supertrace. Following Refs.~\cite{Tiburzi:2004mv,Detmold:2005pt}, we use
\begin{equation}
  \cQ
  =
  \diag
  \left(
    q_u,q_d,q_s,q_j,q_l,q_r,q_u,q_d,q_s
  \right)
,\end{equation}
along with the condition $q_j + q_l + q_r = 0$.
QCD is recovered in the limit of degenerate valence and sea 
quarks only for the particular choice: $q_u = q_j = \frac{2}{3}$, 
and $q_d = q_s = q_l = q_r = - \frac{1}{3}$. Letting the charges 
be arbitrary, however, enables us to track the flow of charge
in loop diagrams.

The relevant electromagnetic couplings in the chiral Lagrangian with quark-diquark symmetry
are contained in the terms~\cite{Hu:2005gf}
\begin{equation} \label{eq:LTM}
\cL 
= 
\frac{\beta e}{2} 
\Big( 
\fH^\dagger \fH \, \cQ \, \bm{\sigma} \cdot \bm{B}
\Big)
+ 
\frac{Q^\prime e}{2 m_Q}
\Big( 
\fH^\dagger \tilde{\bm{\Sigma}} \cdot \bm{B} \, \fH
\Big)
+
\frac{\beta_{E2} \, e}{2 m_Q \Lambda_\chi}
\Big(
\cH^\dagger \tilde{\bm{\Sigma}}^i \cH \cQ \bm{\sigma}^j \Big) \bm{\nabla}_{\{ i } \bm{E}_{ j \}}
,\end{equation}
where $Q^\prime$ is the charge of the heavy quark
and the $5 \times 5$ matrices $\tilde{\bm{\Sigma}}$ are
\begin{equation}
\tilde{\bm{\Sigma}}_{\mu \nu} = 
\begin{pmatrix}
\bm{\sigma}_{\a\b} & 0             \\
                0  & - 2 \, \bm{\cT}_{jk} \\ 
\end{pmatrix}
.\end{equation}
The first term in Eq.~(\ref{eq:LTM}) gives the leading contribution to the decay rate
in the $1/m_Q$ expansion. The contribution of this term at tree level is similar
to the light quark magnetic moment contribution to the transition in
the non-relativistic quark model, i.e. the quark model  reproduces the tree-level
result of HH$\chi$PT upon the identification $1/m_q \to \beta$. The second term in Eq.~(\ref{eq:LTM}) 
is the contribution from the magnetic moment of the heavy quark which is $O(1/m_Q)$. The third term
in Eq.~(\ref{eq:LTM}) is also $O(1/m_Q)$ suppressed. The notation $\{ i j \}$ denotes
symmetrization and tracelessness in the indices $i j$. In the heavy meson sector,
this term contributes to the electric quadrupole moment of the $\bm P^*$ but not to 
the $\bm P^* \to P \gamma$ transition. It does give a contribution to the decay 
$\Xi^* \to \Xi \gamma$, which however is suppressed by $1/m_Q^2 \L_\chi^2$ rather than $1/m_Q \L_\chi$
because it does not interfere with the leading-order contribution. 
Explicitly the formula for the decay width is given by
\begin{equation}
\Gamma ( \Xi^*_a \to \Xi_a \gamma ) = 
\frac{4 \alpha}{9} 
\left[
\left(
\b \cQ_a  - \frac{Q^\prime}{2 m_Q}
+ 
\d \b_a 
\right)^2
+
\frac{3}{4} E_\gamma^2
\left( 
\frac{\b_{E2} \cQ_a }{m_Q \L_\chi}
+
\d \b_{E2} {}_a
\right)^2
\right]
\frac{M_\Xi}{M_{\Xi^*}} E_\gamma^3
,\end{equation}
where $\d\b_a$ and $\d \b_{E2}{}_a$ represent chiral corrections
to the tree-level magnetic dipole and electric quadrupole transition
moments, respectively.
The loop diagrams shown in Fig.~\ref{fig:NLOtrans}
 give the leading non-analytic chiral corrections.

The Lagrangian in Eq.~\eqref{eq:LTM} is easily extended   to \PQCPT.
Using the definition of the super-field $\fH$ in Eq.~\eqref{eq:H}, 
we can expand the Lagrangian in Eq.~\eqref{eq:LTM} to find the doubly 
heavy baryon part and thereby determine the contribution
to the transition moments at tree level. As this contribution only involves
valence quarks, the result is the same as in \CPT~\cite{Hu:2005gf}. 
The coefficients $\b$ and $\b_{E2}$ have the same value as in \CPT\ which 
can be demonstrated by matching to \PQCPT\ in the sea sector. 
In the heavy quark limit, the operator proportional to $\beta_{E2}$ can be neglected.
When $O(1/m_Q)$ hyperfine splittings are included in the loop diagrams, there are non-analytic 
corrections to both $\beta$ and $\beta_{E2}$.
At one-loop order in \PQCPT, we must calculate the diagrams shown in Fig.~\ref{fig:NLOtrans}. 
Explicitly we find
\begin{eqnarray}\label{db}
\delta \beta_a
&=&  
 - \frac{g^2}{24 \pi^2 f^2}
\sum_{b = j, l, r} (\cQ_b - \cQ_a) 
\int_0^1 dx 
\Bigg\{ 
\Delta_1(x) \log \frac{m_{ab}^2}{\mu^2}
- 
m_{ab} R\left(\frac{\Delta_1(x)}{m_{ab}} \right)
\notag \\
&& \phantom{spacerer} +
5 
\left[ \Delta_2(x) \log \frac{m_{ab}^2}{\mu^2}
- 
m_{ab} R\left( \frac{ \Delta_2(x) }{m_{ab}} \right)
\right]
\Bigg\}
,\end{eqnarray}
\begin{eqnarray}\label{db2}
\delta \b_{E2} {}_a
&=&  
\frac{g^2}{24 \pi^2 f^2}
\sum_{b = j, l, r} (\cQ_b - \cQ_a) 
\int_0^1 dx \, x (1-x)
\Bigg\{ 
\frac{\Delta_1(x) m_{ab}}{[\Delta_1(x)]^2 - m_{ab}^2}
R\left( \frac{ \Delta_1(x)  }{m_{ab}} \right)
\notag \\
&& \phantom{spacerer} - 
\frac{ \Delta_2(x) m_{ab}}{ [\Delta_2(x)]^2 - m_{ab}^2}
R\left( \frac{\Delta_2(x)}{m_{ab}} \right)
\Bigg\}
,\end{eqnarray}
In expressing the transition moments, we have made the following 
definitions
\begin{eqnarray}
\Delta_1(x)  &=& -\Delta^*_{ab} + x \Delta^*_{aa} \\
\Delta_2(x)  &=& -\Delta_{ab} + x \Delta^*_{aa}
,\end{eqnarray}
and used the function $R(x)$ defined in Eq.~\eqref{eq:R}.
The \CPT\ result can be deduced by replacing the sea charges
and sea masses with the values of their valence counterparts. 

Ref.~\cite{Hu:2005gf} considered the  loop corrections in 
\CPT\ in the approximation where $\D_H = 0$ and $\sigma = 0$. This
corresponds to taking the heavy quark limit and keeping only
$O(\sqrt{m_q})$ nonanalytic chiral corrections. In this limit
the result can be inferred from the calculation of Ref.~\cite{Amundson:1992yp,Savage:2001jw} for heavy mesons:
\bea \label{approx}
\delta \beta_a =  \frac{g^2}{4 \pi f^2}\sum_b (\cQ_b - \cQ_a) \, m_{ab} 
\qquad \qquad \delta \b_{E2} {}_a = 0 \, .
\eea
For  $\chi$PT with physical parameters, 
using Eq.~(\ref{db}) instead of Eq.~(\ref{approx}) has little effect on the transition
magnetic moment, $\b \cQ_a  - {Q^\prime}/(2 m_Q) + 
\d \b_a$. The transition magnetic moment changes by -14\% (-1\%) for $\mu =500 \, \texttt{MeV}$ ($1500 \,\texttt{MeV}$)
for $a=1$ and by -1\% (+3\%) for  $\mu =500 \,\texttt{MeV}$ ($1500 \,\texttt{MeV}$) for $a=2$.
We conclude that keeping only the $O(\sqrt{m_q})$ nonanalytic chiral correction
is an adequate approximation for estimating the electromagnetic decay width of the $\Xi^*$ in QCD. 
Away from the physical value of the quark mass, chiral corrections become increasingly
important and hence formulae in Eqs.~\eqref{db} and \eqref{db2} should be considered
for lattice extrapolations.

The $x$-integrals can be evaluated analytically as shown in the Appendix. 
In actual lattice calculations of transition moments, 
the current insertion method
is limited by the discrete lattice momenta available. 
The matrix elements contributing to the electromagnetic decay are zero 
 for vanishing photon three-momentum $\bm{q}$. If we wish to extract these matrix elements
 from a lattice simulation, it is necessary to know the 
momentum transfer dependence of the corresponding form factors in order
to perform a model independent zero-momentum extrapolation. 
Formulae for the multipole transition form factors 
can be recovered from our 
expressions above by merely using the replacement
\begin{equation}
m_\phi \to \sqrt{m_\phi^2 - x (1-x) q^2}
,\end{equation}
where $q^2$ is the virtuality of the photon.

\begin{figure}
  \centering
  \includegraphics[width=0.5\textwidth]{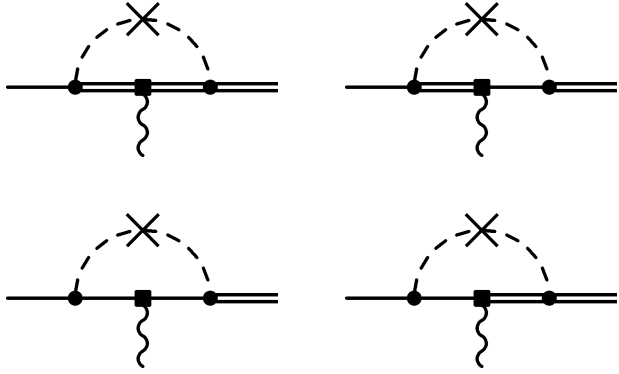}%
  \caption{
    Additional loop graphs for the doubly heavy baryon electromagnetic transitions in \QCPT.  
    The single lines correspond to $\Xi$ baryons, double lines to $\Xi^*$ baryons.
    The dashed lines correspond to mesons, while the wiggly lines correspond to photons.  
    The filled circles denote the non-singlet axial coupling given in Eq.~(\ref{eq:LQ}), 
    the photon coupling is proportional to the magnetic couplings in Eq.~\eqref{eq:LTM}, 
    and crosses denote the quenched hairpin interaction.}
  \label{fig:NLOtransQ}
\end{figure}

While we do not wish to encourage quenched calculations of the 
transition moments, we shall comment briefly on the oddities of the quenched chiral 
calculation at one-loop order. None of the diagrams in Fig.~\ref{fig:NLOtrans}
survive quenching. There is, however, quark mass dependence
at this order, but it arises from the hairpin interactions depicted in 
Fig.~\ref{fig:NLOtransQ}. These diagrams are enhanced over their
\PQCPT\ and \CPT\ counterparts due to the size of the singlet parameter 
$\mu_0^2$. This was first observed for baryon magnetic moments in Ref.~\cite{Savage:2001dy}, 
and applied to baryonic transition moments in Ref.~\cite{Arndt:2003vd}. 
Thus in the chiral limit, the \QCPT\ transition moments have chiral singularities
\begin{eqnarray}
\b_{M1}^Q &\sim& \beta_{M1}^{(0)} {}^Q - \frac{Q^\prime }{m_Q} + \a_{M1} \, \mu_0^2 \, \log m_q + \ldots 
\notag \\
\b_{E2}^Q &\sim& \beta_{E2}^{(0)} {}^Q + \a_{E2} \, \mu_0^2 \, \log m_q + \ldots 
\notag
,\end{eqnarray}
which should be contrasted with the chiral limit behavior in 
both \PQCPT\ and \CPT\
\begin{eqnarray}
\b_{M1} &\sim& \beta_{M1}^{(0)} - \frac{Q^\prime}{m_Q} + \b_{M1}^{(1/2)} \, \sqrt{m_q} + \ldots
\notag \\
\b_{E2} &\sim& \beta_{E2}^{(0)} + \b_{E2}^{(1)} \, m_q \log m_q + \ldots
\notag
,\end{eqnarray}
which are non-singular.

The inclusion of dynamical quarks thus makes a dramatic difference in the chiral behavior of these 
observables. There are two types of dynamical contributions: those from the gauge configurations
and those from the current operator self-contraction. The latter contributions are notoriously difficult
to determine from the lattice. The effective theory can remedy this situation for the electromagnetic current in 
three-flavor simulations. Because we require $\str \cQ = 0$, the choice $q_j = q_l = q_r = 0$ is not 
excluded, i.e.~one can ignore the self-contraction of the current. In this case, one is not determining
the physical transition moments, but rather the physical parameters $\b$ and $\b_{E2}$ in the effective theory.
These in turn can be used to make physical predictions.

\section{Summary} \label{summy}

In this paper, we have extended the chiral Lagrangian with heavy quark-diquark 
symmetry to quenched and partially quenched theories. This extension
allows for the derivation of extrapolation formulae necessary to 
connect lattice QCD data to the chiral limit.
These formulae are easily modified to include the effects of
finite volume, see Ref.~\cite{Arndt:2004bg}. Furthermore light-quark discretization 
effects can be incorporated, but depend on the particular fermion discretization 
employed in the lattice calculation.

As an example, we determined the quark mass dependence of 
doubly heavy baryons in both quenched and partially quenched chiral perturbation theory.
We demonstrated that the hyperfine splitting between $\Xi^*$ and $\Xi$ baryons
is sensitive to chiral non-analytic corrections both in the quenched and partially quenched
theories. Thus careful chiral extrapolation is required to connect lattice data
to the doubly heavy baryon spectrum in the chiral regime.
Surprisingly, we found that the heavy quark-diquark symmetry
relation is rather insensitive to chiral corrections. Despite the possible 
enhancement by factors of $\mu_0^2$ multiplying quenched chiral logarithms, 
this conclusion is true in the quenched theory.

We also investigated the chiral corrections to the electromagnetic transition
moments. The width of the $\Xi^*$ baryons should be dominated by the 
decay  $\Xi^* \to \Xi \gamma$.
The lattice can be used to predict the electromagnetic decay widths
and our results should be helpful for chiral extrapolations in these calculations.
Since the doubly heavy baryon hyperfine splitting is expected to be $ \approx 100 \, \texttt{MeV}$,
lattice pion masses will not allow on-shell intermediate states as one nears 
the chiral regime. This is in contrast to the $\D \to N \gamma$ transitions where
the pion mass must be $m_\pi \gtrsim 300 \, \texttt{MeV}$ to avoid
the multi-particle continuum.
Furthermore, lattice data can test 
heavy quark-diquark symmetry 
relations between the $\Xi^* \to \Xi \gamma$ 
and $P^* \to P \gamma$ decays.

Investigation of heavy mesons and doubly heavy baryons on the lattice provides a way to
explore the implications of heavy quark-diquark symmetry.
By varying the light quark mass, one will be able to see how well
the predictions hold up against chiral corrections. Additionally 
by varying the heavy quark mass, one will be able to see in what
regime this symmetry of the strong interaction  emerges.

\acknowledgments

B.C.T.~would like to thank the Institute for Nuclear Theory 
at the University of Washington for hospitality and 
partial support during the course of this project.
This work was supported by the U.S. Department of
Energy under Grants No.~DE-FG02-05ER41368 (T.M. and B.C.T.), 
DE-FG02-05ER41376, and DE-AC05-84ER40150 (T.M.).

\appendix

\section*{Appendix}

Here we give some formulae which are useful for evaluating the 
$x$-integrals in Eqs.~(\ref{db}) and (\ref{db2}). A useful observation
is that 
\bea
\frac{d R(x)}{d x} =  \frac{x }{x^2-1} R(x) -2  \, .
\eea
Using this it is straightforward to verify the following anti-derivatives: 
\bea\label{ints}
\int dx \, R(x) &=& \frac{1}{8} \frac{1}{x^2-1} R^2(x) + \frac{x }{2} R(x) +
\frac{x^2}{2} \, , \nn \\
\int dx \,\frac{1}{x^2 -1} R(x) &=& -\frac{1}{4} \frac{1}{x^2-1} R^2(x) \, , \nn \\
\int dx \,\frac{x }{x^2 -1} \,R(x) &=& R(x) + 2 x \, , \nn \\
\int dx \,\frac{x^3 }{x^2 -1} \, R(x) &=& \frac{x^2+2}{3} R(x)
+\frac{2 x^3+12 x}{9} \, . 
\eea
All integrals in Eqs.(\ref{db}) and (\ref{db2}) can be evaluated using these identities.
For example, for one of the integrals appearing in  $\delta \beta_a$, we find
\bea
\int_0^1 dx \, 
\left[ 
\Delta_1(x) \log \left(\frac{m_{ab}^2}{\mu^2}\right) \right.
\!\!&-&\!\! \left.
m_{ab} R\left( \frac{\Delta_1(x)}{m_{ab}} \right)
\right]  \nn \\
&=& 
\frac{m_{ab}^2}{\D_{aa}^*}
\int_{-\frac{\Delta_{ab}^*}{m_{ab}}}^{\frac{\Delta_{aa}^* -\Delta_{ab}^*}{m_{ab}}}
du \left[ u \log \left(\frac{m_{ab}^2}{\mu^2}\right) - R(u)\right] \nn \\
&\equiv&
\frac{m_{ab}^2}{\D_{aa}^*}
\left[G\left(\frac{\Delta_{aa}^*-\Delta_{ab}^*}{m_{ab}},\frac{m_{ab}}{\mu}\right)
- G\left(-\frac{\Delta_{ab}^*}{m_{ab}},\frac{m_{ab}}{\mu}\right) \right] \, 
, \nn
 \eea
where
\bea
G(x,y) &=& x^2 \log(y) - \frac{x^2}{2}  
-\frac{1}{8} \log^2 \left( 
\frac{x - \sqrt{x^2 - 1 + i \epsilon}}{x + \sqrt{x^2
- 1 + i \epsilon}} \right) \nn \\
&&- \frac{x\sqrt{x^2-1}}{2} \log \left( 
\frac{x - \sqrt{x^2 - 1 + i \epsilon}}{x + \sqrt{x^2
- 1 + i \epsilon}} \right) 
 \, . \nn
\eea

\bibliography{hb}

\end{document}